\documentclass[10pt]{iopart} 
\usepackage{graphicx}
\usepackage[dvipsnames]{xcolor}
\usepackage{cite}
\begin{document}

\title[]{New Optical Models for the Accurate Description of the Electrical Permittivity in Direct and Indirect Semiconductors}

\author{K. Lizárraga$^{*1}$, L. A. Enrique-Morán$^*$, A. Tejada$^{*,\dagger}$, M. Piñeiro$^*$, P. Llontop$^*$, E. Serquen$^*$, E. Perez$^*$, L. Korte$^\dagger$, and J. A. Guerra$^{*,\dagger,2}$}

\address{$^{*}$Departamento de Ciencias, Sección Física, Pontificia Universidad Católica del Perú, Av. Universitaria 1801, Lima 32, Peru}
\address{$^{\dagger}$Helmholtz-Zentrum Berlin für Materialien und Energie GmbH, Division Solar Energy, Kekuléstraße 5, 12489 Berlin, Germany}
\ead{$^1$kevin.lizarraga@pucp.edu.pe, $^2$guerra.jorgea@pucp.edu.pe}
\vspace{10pt}
\begin{indented}
\item[]June 2022
\end{indented}

\begin{abstract}
We propose new models to describe the imaginary part of the electrical permittivity of dielectric and semiconductor materials in the fundamental absorption region. We work out our procedure based on the well-known structure of the Tauc-Lorentz model and the band-fluctuations approach to derive a 5-parameter formula that describes the Urbach, Tauc and high-absorption regions of direct and indirect semiconductors. Main features of the models are the self-consistent generation of the exponential Urbach tail below the bandgap and the incorporation of the Lorentz oscillator behaviour due to electronic transitions above the fundamental region. We apply and test our models on optical data of direct (MAPbI$_{3}$, GaAs and InP), indirect (GaP and c-Si), and amorphous (a-Si) semiconductors, accurately describing the spectra of the imaginary part of the electrical permittivity. Lastly, we compare our models with other similarly inspired models to assess the optical bandgap, Urbach tail and oscillator central resonance energy.
\end{abstract}

%
\vspace{2pc}

\submitto{\JPD}
%
\maketitle
 
\ioptwocol

\section{Introduction}
Models describing the fundamental absorption and high absorption regions of dielectric non-excitonic materials are scarce, in particular there is no equivalent Tauc-Lorentz model for direct electronic transitions materials. The correct determination of quantities such as the optical bandgap, Urbach energy and oscillator central energies, relies on their physical validity. The knowledge of the optical bandgap and Urbach energy are essential for photoelectric devices design \cite{guerra,tejada,ugur}. In addition, the oscillator's central energy is fundamental for exploring distinct electroniv band transitions and material simulations \cite{shirayama}.

The absorption edge or fundamental absorption region, where the onset of band-to-band absorption takes place, is typically overlapped with disorder induced localized states, i.e. Urbach tails, whereas the high absorption regime is governed by band-to-band transitions which can be properly described by the classical Lorentz oscillator model and its variants \cite{franta}. In the last two decades, these regions have been described by the models of Tauc \cite{tauc}, Jellison and Modine \cite{modine}, Ferlauto \cite{ferlauto}, and more recently by Ullrich \cite{ullrich,ullrich2}, O'Leary \cite{oleary,thevaril} and our group \cite{guerra}.

Whereas most of these models have found their way to commercial software and are widely used to model the properties of distinct materials, the most commonly used model is perhaps the Tauc-Lorentz (TL) model. Although Tauc's parabolic absorption spectra shape describes the absorption edge of indirect and amorphous materials, nowadays is used for direct semiconductors as well when fitting optical transmittance, reflectance and ellipsometric data, underestimating the optical bandgap \cite{fujiwara}. Additionally, the TL model does not take into account Urbach tails, attributed to disorder-induced localized states and thermal effects. Thus, when fitting optical data of direct electronic transitions materials, besides the difference in shape in the fundamental absorption, the exponential Urbach tail is actually fitted with the parabolic shape of Tauc's model. This is the main problem of using the TL model for retrieving the optical bandgap. 

Ferlauto et al.\cite{ferlauto} developed a model by incorporating an exponential tail to the TL model. This model came to be known as Cody-Lorentz (CL) model. Ullrich et al \cite{ullrich,ullrich2} modelled the absorption coefficient of direct electronic transitions materials through the addition of an exponential behaviour below the bandgap that satisfies the first derivative continuity condition. O'Leary et al \cite{oleary,thevaril} modelled the absorption coefficient of amorphous silicon by incorporating an exponential tail to the valence-band density of states, which was further convoluted with the conduction-band density of states to calculate the fundamental absorption. In a previous work \cite{guerra}, we modified the absorption edge by introducing band-fluctuations to the joint density of states for both direct and indirect electronic transitions materials. In this way Urbach tails are incorporated in both types of models \cite{guerra}. These models were then used to analyze experimental data.

In order to arrive at an accurate description of the optical absorption, we propose in this work to apply the procedure of Tauc-Lorentz to the Ullrich, O'Leary and bands-fluctuations approaches, thus, unifying the absorption edge and high absorption regions in a single equation for each model.

We proceed and develop each of the aforementioned models in a single electronic Joint Density of States (JDOS) which is proportional to the optical absorption coefficient. We extend the models to include a Lorentz oscillator component for the high absorption region. We compare the extended models with experimental data for direct electronic transitions materials such as methylammonium lead iodide (MAPI), a metal halide pervoskite whose bandgap can be controlled stoichiometrically. Property that is exploited e.g. for tandem solar cells when paired with silicon \cite{lars1,menzel}; gallium arsenide (GaAs) and indium phosphide (InP) whose applications are in high-speed, optoelectronic and photovoltaic devices \cite{chennupati,waferworld}; indirect electronic transitions materials such as gallium phosphide (GaP) which is used typically in Light Emitting Devices (LED) technology \cite{vaclavik}; crystalline silicon (c-Si), which is widely used in electronic and photovoltaic applications \cite{kearns,ozevin}; and amorphous silicon (a-Si), which is used in thin film solar cells \cite{guha,slaoui}.

\section{Established absorption edge models}
Here we summarize current models for the absorption edge. We extend the Tauc-Lorentz approach to these models to account for higher energy band-to-band transitions and deliver analytical expressions for each model. 

\subsection{Fundamental Absorption}
\begin{figure*}[htb]\centering 
\includegraphics[scale=0.45]{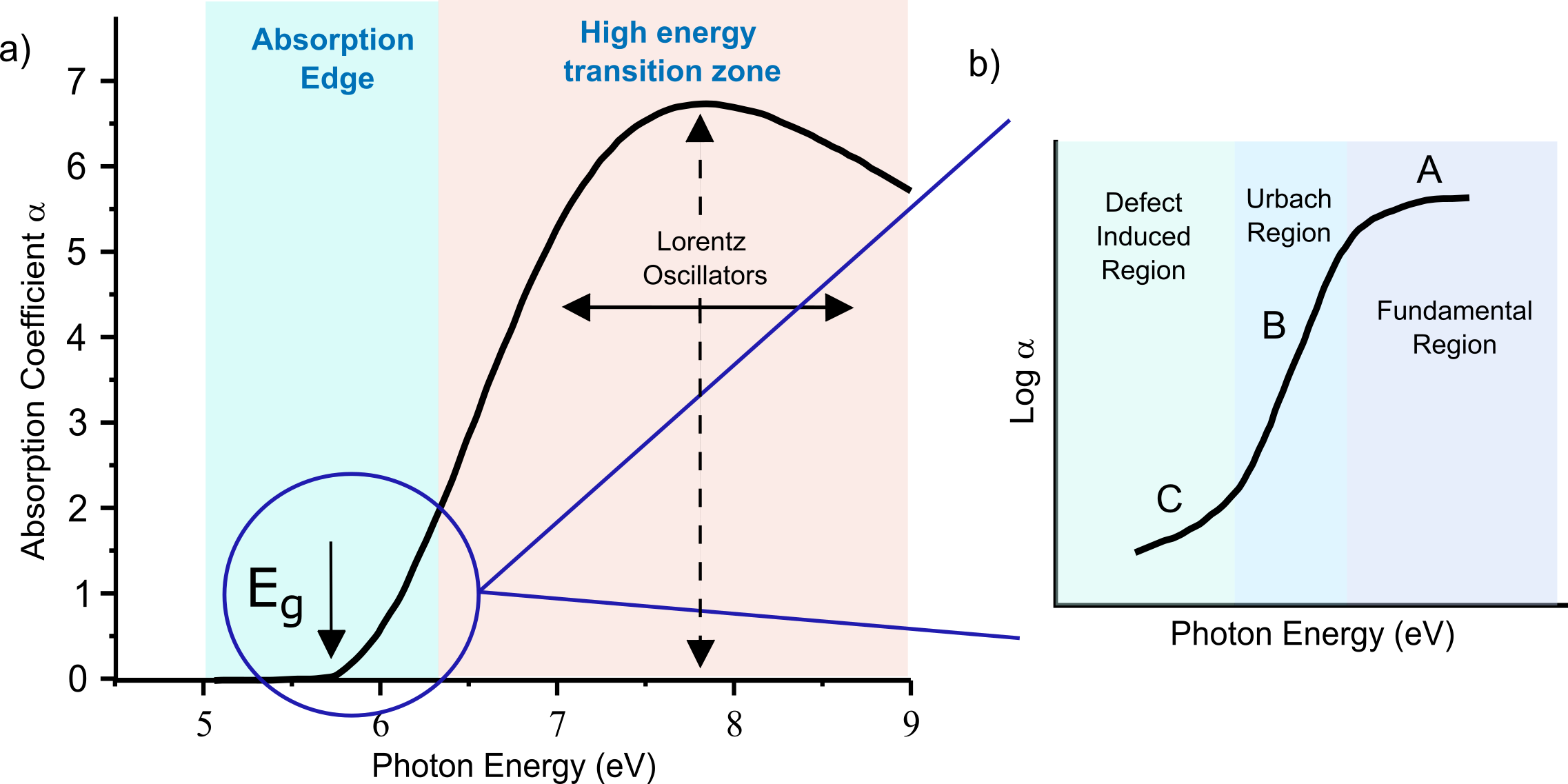}
\caption{Schematic of the absorption  coefficient ($\alpha$) near the absorption edge and high energy region (a). The different zones of the absorption edge region are illustrated in logarithmic scale in figure (b). Here, region A corresponds to the fundamental absorption, region B to the Urbach tail and region C to the absorption induced by impurities present in the material.}
\label{fig:fund_reg}.
\end{figure*}
The description of the absorption coefficient near the absorption edge is typically described in three zones as depicted in figure \ref{fig:fund_reg}. Region A corresponds to the parabolic band approximation, usually fitted with Tauc's equation, whose Region B corresponds to the universally observed Urbach tail associated with disorder-induced localized states. Region C is associated to  defect-induced localized states. The latter zone is typically studied by optical absorption measurements of bulk crystals \cite{weingartner,weingartner2,bickermann}, or thin films by UV-Excited Photoelectron Spectroscopy (UPS) on states near the valence band edge and can be modeled independently \cite{menzel2,menzel}. For our  purposes this region won't be part of our analysis.

The description of the optical properties of semiconductors was developed in the early 60s. It is based on the existence of long-range order, electron k-vector conservation and Fermi's golden rule. The absorption coefficient is proportional to the electronic transition rate, which in the most general case can be written as \cite{cardona}:
\begin{equation}
    R_{cv}=R\sum_{k_{c},k_{v}} |M_{cv}|^{2}\delta(E_{c}-E_{v}-\hbar\omega \pm E_{\Omega})\delta_{k_{c},k_{v}+q\pm k_{\Omega}}. \label{Rfirst}
\end{equation}
Here $R$ is $2\pi/\hbar(\vec{E}e/2\omega m_{e})^{2}$. $E_{c}$ and $E_{v}$ are the electron conduction and valence band energy states. $|M_{cv}|$ is the electronic transition matrix element whose behaviour is typically taken constant near the absorption edge, whilst is modeled by means of the Lorentz oscillator for higher energies. The general electronic transition involves a photon and a phonon with energies $\hbar\omega$ and $E_{\Omega}$, and momentum $q$ and $k_{\Omega}$, respectively.  Thus, we have the energy conservation term $E_{c}=E_{v}+\hbar\omega\mp E_{\Omega}$, and the momentum conservation $k_{c}=k_{v}+q\pm k_{\Omega}$. For $E_{\Omega}\ll \hbar\omega$ and $q\ll 1$, the terms $q$ and $E_{\Omega}$ are neglected. For the case of direct electronic transitions materials, the most probable transitions keeps $k_{c}=k_{v}$, whilst in the case of indirect transitions we have $k_{c}=k_{v}\pm k_{\Omega}$. In the framework of the effective mass approximation (electronic bands parabolic approximation), the conduction ($D_{c}$) and valence $D_{v}$ bands electronic density of sates are \cite{jackson}, i.e.,
\begin{equation}
D_{c}(E_{c})=\frac{\sqrt{2}m_{e}^{*3/2}}{\pi^{2}\hbar^{3}} (E_{c}-E_{g})^{1/2} \Theta(E_{c}-E_{g}), \nonumber
\end{equation}
\begin{equation}
D_{v}(E_{v})=\frac{\sqrt{2}m_{h}^{*3/2}}{\pi^{2}\hbar^{3}} (-E_{v})^{1/2} \Theta(-E_{v}). \label{dos.ind1}
\end{equation}
Here $m_e$ and $m_{h}$ are the electron and hole masses; $E_{c}$ and $E_{v}$ are the conduction and valence band energy; $E_{g}$ is the band gap energy; and, $\Theta$ is the step-function. For the case of direct semiconductors, the transition rate $R^{d}_{cv}$ can be written in terms of the JDOS $D_{cv}$ with $E_{cv}=E_{c}-E_{v}$ as the band energy difference \cite{cardona}.
\begin{equation}
        R^{d}_{cv} = R\int dE_{cv} |M_{cv}|^{2}D_{cv}(E_{cv})\delta(E_{cv}-E).
        \label{dir_conv}
\end{equation}
$D_{cv}(E_{cv})$ displays a square root shape versus $E_{cv}$ and it is proportional to the reduced effective mass $\mu^{*-1}=m_{e}^{*-1}+m_{h}^{*-1}$:
\begin{equation}
    D_{cv}(E_{cv})=\frac{\sqrt{2}\mu^{*3/2}}{\pi^{2}\hbar^{3}}(E_{cv}-E_{g})^{1/2}\Theta(E_{cv}-E_{g}). \label{dos.dir1}
\end{equation}
Consequently, the electronic transition rate is:
\begin{equation}
    R^{d}_{cv}= R|M_{cv}|^{2} \frac{\sqrt{2}\mu^{*3/2}}{\pi^{2}\hbar^{3}}(E-E_{g})^{1/2}\Theta(E-E_{g}). \label{dir_r}
\end{equation}
On the other hand, in the case of indirect electronic transitions materials, all energetically possible transitions between initial valence (v) and final conduction (c) states must be considered. For this reason, $R^{i}_{cv}$ is written in terms of the valence $D_{v}$ and conduction $D_{c}$ electronic Density Of States (DOS), i.e.,
\begin{equation}
R^{i}_{cv}=R|M_{cv}|^{2} \int dE_{cv} J_{cv}(E_{cv}) \delta(E_{cv}-E),
\label{ind_conv}
\end{equation}
with 
\begin{equation}
    J_{cv}(E_{cv})= \frac{2(m_{e}^{*}m_{h}^{*})^{3/2}}{\pi^4 \hbar^6} \frac{\pi}{8} (E_{cv}-E_{g})^2 \Theta(E_{cv}-E_{g})
\end{equation}
as the indirect JDOS and $E_{cv}=E_{c}-E_{v}$. Thus the indirect electronic transition rate is
\begin{equation}
R^{i}_{cv}=R|M_{cv}|^{2} \left( \frac{(m_{e}^*m_{h})^{3/2}}{4 \pi^{3}\hbar^{6}} \right) (E-E_{g})^{2} \Theta(E-E_{g}).
\label{ind_r}
\end{equation}
For the case of amorphous materials, Tauc successfully derived a straightforward formula for the fundamental absorption. He relaxed the conservation of the wave vector $k$, allowing all possible transitions \cite{tauc,cardona}. Coincidentally, the simplification proposed by Tauc has the same quadratic behaviour as for indirect electronic transitions materials, i.e.,
\begin{equation}
\epsilon_{2}\approx \frac{(E-E_{g})^2}{E^2} \label{tauc}
\end{equation}
This analysis was key to understanding the fundamental absorption of a-Si\cite{tauc}. Equations (\ref{dir_r}) and (\ref{ind_r}) are the starting set of equations for the extensions we propose in this work. They can be coupled to Lorentz oscillator theory and fulfill the Kramers-Kroning (KK) condition \cite{modine}, i.e. the function must fall to zero in the infinity. However, these models do not consider the Urbach tail overlap on the fundamental absorption and thus bias the bandgap determination.

Urbach found experimentally in 1953 an exponential behavior in the optical absorption edge of AgBr crystals \cite{urbach}:
\begin{equation}
\epsilon_{2} \approx e^{(E/E_{U})} \label{urbach_tail}.
\end{equation}
Here $E_{U}$ is the width of the tail, known as Urbach energy. The nature of the Urbach rule could be attributed to many factors such as the presence of longitudinal-optical (LO) phonons involved in electronic transitions \cite{kurik}, the exciton-phonon interaction (exciton self-trapping) \cite{Ueta} or by the Franz-Keldysh effect, in which Bloch waves can tunnel in the bandgap region due to an electric field originating from vibrations in the amorphous network \cite{redfield}. Despite theur diverse origins, the current consensus is that thermal effects and static disorder are the main reasons behind their appearance \cite{cody,cody1981,shimakawa}. 

There are various models trying to describe the behavior of tail states, such as the models of Ullrich \cite{ullrich2,ullrich}, O'Leary, Malik \cite{oleary,oleary2,oleary3,malik}, Orapunt \cite{orapunt}, Thevaril \cite{thevaril2,thevaril} and, most recently, by our group \cite{guerra}. Ullrich's and O'Leary's models are based on a modified DOS. In both models, the DOS is a piecewise function of tail states and extended states. Ullrich modifies eq. (\ref{dos.dir1}) as:
\begin{equation}
D^{U}_{cv}(E_{cv})=D_{0}\left\{
                \begin{array}{ll}
                  (E_{cv}-E_{g})^{1/2} \ \ \ \ \ , E_{cv} \geq E_{\mathit{cv_{T}}} \\
                  \frac{1}{\sqrt{2\beta}}e^{\beta(E_{cv}-E_{cv_{T}})}  \  , E_{cv} < E_{\mathit{cv_{T}}} 
                \end{array} 
                   \right.    
\end{equation}
for the case of direct electronic transitions materials. $D_{0}$ is an overall constant, $\beta$ is the inverse of Urbach energy and $E_{cv_{T}}=E_{g}+1/(2\beta)$ denotes the point satisfying the first derivative continuity condition for a smooth transition between the exponential tail and the square root describing the transition between extended states. On the other hand, O'Leary proposes eq. (\ref{dos.ind1}) for amorphous Si:
\begin{eqnarray}
D^{O}_{v}(E_{v})=&\frac{\sqrt{2}m_{h}^{*3/2}}{\pi^{2}\hbar^{3}} \nonumber \\ 
&\times \left\{
                \begin{array}{ll}
                  \frac{1}{\sqrt{2\beta_{v}}}e^{\beta_{v}(E_{v_{T}}-E_{v})} &,  E_{v} > E_{v_{T}} \\
                  (-E_{v})^{1/2}  &,  E_{v} \leq E_{v_{T}}             
                \end{array},  
                \right.
\end{eqnarray}
\begin{equation}
D^{O}_{c}(E_{c})=\frac{\sqrt{2}m_{e}^{*3/2}}{\pi^{2}\hbar^{3}} (E_{c}-E_{g})^{1/2} \Theta(E_{c}-E_{g}),             
\end{equation}
where $E_{v_{T}}$ guarantees the continuity of the DOS. Tails are considered in the valence DOS only \cite{oleary}. 

We can calculate the respective electronic transition rates, for direct
\begin{equation}
R^{U}_{cv}= R|M_{cv}|^{2} D^{U}_{cv}(E), \label{r_u}
\end{equation}
and indirect (amorphous) materials
\begin{equation}
R^{O}_{cv}= R|M_{cv}|^{2} J^{O}_{cv}(E), \label{r_o}
\end{equation}
where
\begin{equation}
J^{O}_{cv}(E)=J_{0}\frac{1}{\beta^{2}} j^{O}_{cv}(\beta(E-E_{g}))
\end{equation}
with $J_{0}$ being an overall constant, and
\begin{eqnarray}
j^{O}_{cv}(z)=\left\{
                \begin{array}{ll}
                  \Xi(z)  & ,  z \geq 1/2 \\
                  \frac{1}{\sqrt{2}}e^{\left(z-1/2\right)} Y(0)  & , z < 1/2 \\              
                \end{array},  
                \right. 
\label{jo}                
\end{eqnarray}
with
\begin{equation}
   \Xi(z) = z^{2}\Sigma \left( \frac{z-1/2}{z} \right) +\frac{1}{\sqrt{2}}Y(z-\frac{1}{2})e^{\left( z-\frac{1}{2} \right)},\label{xi}
\end{equation}
and
\begin{eqnarray}
\Sigma(z) = \frac{\pi}{8}+\frac{\sqrt{z-z^2}}{4}&(2z-1) \nonumber\\
&+\frac{\sqrt{z-1}}{4}\frac{\textrm{Sinh}^{-1}(\sqrt{z-1})}{\sqrt{1-z}},
\end{eqnarray}
\begin{equation}
Y(z) = \sqrt{z}e^{-z} + \frac{\sqrt{\pi}}{2}\textrm{Erfc}(\sqrt{z}).
\end{equation}

Eqs. (\ref{r_u}) and (\ref{r_o}) can describe the whole fundamental region smoothly. Urbach tails can be generated by means of  band-fluctuations. This approach has been successfully applied to amorphous Si:H, SiC:H, SiN, crystalline GaAs and nano-crystalline GaMnN \cite{guerra, guerra2}, nano-crystalline methylammonium lead iodide \cite{guerra3}, and formamidinium cesium lead mixed-halide \cite{tejada}. Band-fluctuations give rise to Urbach tails. Thus, the determined optical bandgap, by fitting these models, is free of bias. Details on this approach can be found elsewhere \cite{guerra}. 
According to the band-fluctuations  model, the direct and indirect JDOS are expressed as:
\begin{equation}
D_{cv}^{G}(E) =-\frac{\sqrt{2}\mu^{*3/2}}{\pi^{2}\hbar^{3}} \frac{1}{2}\sqrt{\frac{\pi}{\beta}}\textrm{Li}_{1/2} \left( -e^{\beta(E-E_{g})} \right), \label{D_g}
\end{equation}
\begin{equation}
J_{cv}^{G}(E)=-\frac{2(m_{e}^{*}m_{h}^{*})^{3/2}}{\pi^{4}\hbar^{6}} \frac{1}{4}\frac{\pi}{\beta^{2}}\textrm{Li}_{2} \left( -e^{\beta(E-E_{g})} \right) \label{I_g}.
\end{equation}

Eqs. (\ref{D_g}) and (\ref{I_g}) describe an exponential Urbach tail and the square-root$/$quadratic behavior for direct$/$indirect (amorphous) semiconductors in the limiting cases below and above the bandgap, respectively. 

\subsection{High Energies Transition Zone}
Above the fundamental absorption, band-to-band electronic transitions are characteristic and can be described by the Lorentz model or driven harmonic oscillator (DHO). The solid is considered classically by the assumptions that electrons are bounded to nuclei harmonically and with dissipative effects \cite{fujiwara,mitalmog}. The real ($\varepsilon_{1}$) and imaginary ($\varepsilon_{2}$) parts of the dielectric constant are:
\begin{equation}
\varepsilon_{1} =1+\frac{A(E_{c}^{2}-E^{2})^{2}}{(E_{c}^{2}-E^{2})^{2}+B^{2}E^{2}}\end{equation}
\begin{equation}
\varepsilon_{2} =\frac{ABE}{(E_{c}^{2}-E^{2})^{2}+B^{2}E^{2}}. \label{im.lor}
\end{equation}

Equation (\ref{im.lor}) is the so called Lorentz term and is parametrized by the height $A$, width $B$ and central energy $E_{c}$ of the oscillator peak. The presence of a single peak is rather uncommon due to the many-body system of a crystalline solid. For this reason, the common agreed extension is to add several oscillators as
\begin{equation}
\varepsilon_{2} =\sum_{i}\frac{A_{i}B_{i}E}{(E_{ci}^{2}-E^{2})^{2}+B_{i}^{2}E^{2}}.
\end{equation}

\subsubsection{Tauc-Lorentz Model.}\ 
Attempts to find a modified Lorentz model that includes Tauc's law or Urbach tail are many. The first of such model was proposed by Forouhi and Bloomer \cite{forouhi}. This model lacks of time reversal symmetry and the integral in the sum rule diverges \cite{franta}. However, it has served as an inspiration of forthcoming models.

Subsequent models overcoming above issues were proposed by Campi and Coriasso (CC) \cite{campi}, and then by Jellison and Modine (JM) \cite{modine}. Despite both models describing the driven harmonic oscillator coupled with Tauc's law, the most popular is the JM model and is usually cited as the Tauc-Lorentz (TL) model. It has been implemented in the majority of optical analysis softwares nowadays.

The idea behind the TL model is to modify the energy independent transition matrix element $M_{cv}$ present in eq. (\ref{ind_r}) to an energy dependent function (the Lorentz oscillator). This is achieved by multiplying the imaginary dielectric constant of the Lorentz model, eq. (\ref{im.lor}), with the quadratic behavior, eq. (\ref{ind_r}), for indirect semiconductors, i.e. for energies above $E_{g}$,
\begin{equation}
\varepsilon^{\mathit{TL}}_{2}(E) =\varepsilon^{T}_{2}(E) \times \varepsilon^{L}_{2}(E), \label{tl_model}
\end{equation}
\begin{equation}
\varepsilon^{\mathit{TL}}_{2}(E) = \frac{(E-E_{g})^{2}}{E^{2}}\frac{ABE \times \Theta(E-E_{g}) }{(E^{2}-E_{c}^{2})^{2}+B^{2}E^{2}} \label{tl_2_model}.
\end{equation}  
This model is physically consistent, it is Kramers-Kroning consistent, and it converges into the classical Lorentz model for high energies \cite{franta2013,franta}. Despite the success of TL model, it presents two main problems: (i) It underestimates the bandgap by not including the Urbach tail. (ii) Its shape is for indirect (and amorphous) electronic transitions materials only. Since the functional behaviour arises from the quadratic dependence with photon energy, in practice, the Urbach tails end being modeled by the parabolic shape of the Tauc model, biasing the determination of the bandgap. 

We aim our attention to the different models that can be obtained by emulating the approach of eq. (\ref{tl_model}), referring to them as modified TL models.

\subsection{Modified Tauc-Lorentz Models}
The first modified TL model was proposed by Ferlauto \cite{ferlauto}. It was later known as the Cody-Lorentz (CL) model to honor Cody's research on a-Si. More recently Franta et al \cite{franta,franta2013} proposed two new models. Their idea was based on replacing the DHO with the Lorentz function. Likewise the CL model, the latter yielded an analytical $\varepsilon_{1}$ after Kramers-Kroning transformation. Here, we extend the spectral dependence of Ullrich, O'Leary and BF by incorporating the Lorentz oscillator.

\subsubsection{Cody-Lorentz Model.}\ 
The absence of the Urbach tail in the TL model was tackled by Ferlauto et al. He incorporated the exponential behavior of Urbach tails to the TL model through a piece-wise function \cite{ferlauto}, i.e.,
\begin{equation}
\varepsilon_{2}(E)=\left\{
                \begin{array}{ll}
                  \frac{E_{1}}{E}  e^{\left( E-E_{t}/E_{U} \right)}  & ,  0 < E \leq E_{t} \\
                  G(E)\frac{ABE}{(E^{2}-E_{c}^{2})^{2}+B^{2}E^{2}}  & ,  E > E_{t} \\              
                \end{array}  
                \right.  \\
\end{equation}
Here $E_{t}$ is a fitting parameter that defines the transition energy from the exponential behavior to the TL one. $G(E)$ can be either $G_{T}(E)=(E-E_{g})^{2}/E^{2}$ (which correspond to a constant momentum matrix element) or the function proposed by Ferlauto to fit a-Si:H: \cite{ferlauto}:
\begin{equation}
G_{C}(E)=\frac{(E-E_{g})^{2}}{(E-E_{g})^{2}+E_{p}^{2}}. \label{cody_model}
\end{equation}
$E_{p}$ represents a second-order transition before Lorentz oscillator takes the shape. It is important to remark that this equation arises from the consideration of a constant dipole matrix element. And, even though it may seem the natural extension of the TL model, it carries two main problems, i.e. the discontinuity produced by $E_{t}$ for the first derivative of $\varepsilon_{2}$ and the quadratic dependence of $G(E)$ excluding direct semiconductor's absorption behaviour. 

\subsubsection{Ullrich-Lorentz (UL) Model.}\ 
The problems of the CL model can be overcomed by Ullrich's work for direct electronic transitions materials. Following the aforementioned procedure of Jellison-Modine model, we multiply the Lorentz dielectric constant, eq. (\ref{im.lor}), to the Ullrich's continuous dielectric function derived from eq. (\ref{r_u}), i.e.,
\begin{equation*}
\varepsilon^{\mathit{UL}}_{2}(E) =\varepsilon^{U}_{2}(E) \times  \varepsilon^{L}_{2}(E)
\end{equation*}
\begin{eqnarray}
\varepsilon^{\mathit{UL}}_{2}(E) = &\frac{ 4\pi^{2}e^{2} |M_{cv}^{2}|D_{cv}^{U}(E)}{n m_{e}^{2}E^{2}} \nonumber\\
& \times \frac{ABE}{(E^{2}-E_{c}^{2})^{2}+B^{2}E^{2}}\Theta(E-E_{g}) \label{ul_2_model},
\end{eqnarray}  
\begin{eqnarray}
\varepsilon^{\mathit{UL}}_{2}(E)=&C \frac{L(E)}{E\sqrt{\beta}} \nonumber\\
&\times \left\{
                \begin{array}{ll}
                  (\beta(E-E_{g}))^{1/2} & , \ E \geq E_{g}+\frac{1}{2\beta} \\
                  \frac{1}{\sqrt{2}}e^{\beta(E-E_{g})}   & , \ E < E_{g}+\frac{1}{2\beta} \\              
                \end{array} . 
                \right. \label{ul-model} 
\end{eqnarray}

Here $C$ is a constant equal to $4\pi^{2}e^{2} |M_{cv}^{2}|D_{0}/n m_{e}^{2}$, where $n$ is the refractive index. $L(E)$ is the Lorentz function, i.e. eq (\ref{im.lor}) divided by $E$, and $\beta$ is the Urbach slope. 

The UL model offers continuity of eq. (\ref{ul-model}) in the first derivative and the square root shape of the absorption coefficient for direct electronic transitions materials as well as the Lorentz behaviour for higher energies.

\subsubsection{O'Leary-Lorentz (OL) Model.} \ 
We now use the same procedure devise by Jellison-Modine for the O'Leary model for indirect/disordered electronic transitions materials, i.e.
\begin{equation*}
\varepsilon^{\mathit{OL}}_{2}(E) =\varepsilon^{O}_{2}(E) \times  \varepsilon^{L}_{2}(E)
\end{equation*}
\begin{eqnarray}
\varepsilon_{2}^{\mathit{OL}}(E)&=\frac{\tilde{C}}{E} \frac{L(E)}{\beta^{2}} \nonumber\\
&\times\left\{
                \begin{array}{ll}
                   \Xi(\beta(E-E_{g}))  &, E \geq E_{g} + \frac{1}{2\beta}  \\
                  \frac{Y(0)}{\sqrt{2}}e^{\left(\beta(E-E_{g})-\frac{1}{2}\right)} &, E < E_{g}+\frac{1}{2\beta} \\              
                \end{array},
                \right.  \label{ol-model}
\end{eqnarray}
with $\tilde{C}$ equal to $4\pi^{2}e^{2} |M_{cv}^{2}|J_{0}/n m_{e}^{2}$. This model is continuous in the first and second derivatives.

\subsubsection{Band-Fluctuations-Lorentz (BFL) Model.}\ 
The UL and OL models are excellent modifications to the Lorentz model since they carry all the information needed for the fundamental absorption of both direct and indirect (amorphous) materials and fulfill the requirements for a Kramers-Kroning transformation. In addition, the band-fluctuations model offers a good description of the fundamental absorption as it was shown in our previous work \cite{guerra}. The extension of the BF following the approach of Jellison-Modine is
\begin{equation*}
\varepsilon^{\mathit{BFL}}_{2}(E) =\varepsilon^{\mathit{BF}}_{2}(E) \times  \varepsilon^{L}_{2}(E),
\end{equation*}
\begin{equation}
\varepsilon^{\mathit{BFL}}_{2,d}(E) =-C\frac{1}{2E}\sqrt{\frac{\pi}{\beta}}\textrm{Li}_{1/2} \left( -e^{\beta(E-E_{g})} \right)L(E), \label{bfl-dir} 
\end{equation}
for direct ($d$) electronic transitions materials, and
\begin{equation}
\varepsilon^{\mathit{BFL}}_{2,i}(E)=-\tilde{C}\frac{\pi}{4\beta^{2}E}\textrm{Li}_{2} \left( -e^{\beta(E-E_{g})} \right)L(E), \label{bfl-ind}
\end{equation}
for indirect/amorphous ($i$) electronic transitions  materials. The BFL model can describe the absorption coefficient near the band edge if either direct and indirect (amorphous) electronic transitions materials from the same principles. The description of Urbach tails in a single equation, the asymptotic behaviour of the polylog function as $\sqrt{E-E_{g}}$ and $(E-E_{g})^2$ for the direct and indirect cases,  respectively, as well as the soft and continuous transition from the fundamental to the high transition zone. 

\subsubsection{Monolog-Lorentz Model.}\ 
The Polylogarithmic functions of order $2$ and $1/2$ that appear in the BF and BFL models are available in most software mathematical analysis environments such as Wolfram Mathematica, MatLab, Python, but not in other more common software for fitting analysis. For this reason, we propose an analytic-handed model based on the band-fluctuations approach. This is done by performing the fluctuations operation on the linear scale (Tauc-scale) of the JDOS for the direct and indirect cases, respectively. Subsequently, the approach of Jellison-Modine is used to obtain a Kramers-Kroning consistent expresion of $\varepsilon_{2}$. For direct electronic transitions materials $\varepsilon_{2}$ is
\begin{equation}
\varepsilon_{2,d}^{\mathit{M}} (E) =  \frac{C}{E^{2}}\sqrt{\frac{1}{\beta}} \textrm{log}^{1/2}\left(1+e^{\beta(E-E_{g})} \right), \label{m-dir}
\end{equation}
and then multiplied by the the Lorentz term of eq. (\ref{im.lor}) it becomes
\begin{equation}
\varepsilon_{2,d}^{\mathit{ML}} (E) =C\frac{1}{E}\sqrt{\frac{1}{\beta}} \textrm{log}^{1/2}\left(1+e^{\beta(E-E_{g})} \right)L(E). \label{ml-dir} 
\end{equation}
For indirect/amorphous materials, $\varepsilon_{2}$ is
\begin{equation}
 \varepsilon_{2,i}^{\mathit{M}} (E) = \frac{\tilde{C}}{E^{2}}\frac{\pi}{8\beta^{2}}\textrm{log}^{2} \left( 1+e^{\beta(E-E_{g})} \right), \label{m-ind}
\end{equation}
and multiplying by the Lorentz it becomes
\begin{equation}
\varepsilon_{2,i}^{\mathit{ML}} (E) =\tilde{C}\frac{\pi}{8\beta^{2}E}\textrm{log}^{2} \left( 1+e^{\beta(E-E_{g})} \right)L(E). \label{ml-ind}
\end{equation}
Despite this model being straightforward based on the observation of the functional behaviour generating an Urbach tail rather than physical principles, it can give bandgap, Urbach and oscillators central energies very close, if not he same, as the retrieved by the BFL models. For instance, the asymptotic behaviour of eq. (\ref{m-dir}) for direct electronic transitions materials is
\begin{equation}
\varepsilon_{2,d}^{\mathit{M}}(E) \approx \frac{C}{E^2} \left\{
                \begin{array}{ll}
                  \sqrt{E-E_{g}}  & , E \gg E_{g} \\
                  \sqrt{\frac{1}{\beta}}e^{\frac{\beta}{2}(E-E_{g})}   & , E \ll E_{g} \\            
                \end{array}  
                \right. 
                \label{m-asymp-dir} 
\end{equation}
and, for indirect, eq. (\ref{m-ind}) behaves asymptotically as
\begin{equation}
\varepsilon_{2,i}^{\mathit{M}}(E) \approx \frac{\tilde{C}}{E^2}\pi               \left\{
                \begin{array}{ll}
                  \frac{1}{8}(E-E_{g})^2  & , E \gg E_{g} \\
                  \frac{1}{8\beta^2} e^{2\beta(E-E_{g})}   & , E \ll E_{g} \\  \end{array}  
                \right. .
                \label{m-asymp-indir} 
\end{equation}
Note how the exponential tail and the square root (parabolic) shape versus the photon energy are recovered for direct (indirect) electronic transitions.  

\section{Dimensionless JDOS formalism}
There are certain mathematical features in the implementation of the aforementioned models that are remarkable. For example, the capability to perform a single fit of the fundamental region and the band-to-band transition zone with 5 parameters only: the constant C, Urbach slope $\beta$, bandgap $E_{g}$, oscillator central energy $E_{c}$ and the oscillator broadening $B$. Furthermore, it gives the advantage of discriminating tail states from band-to-band electronic transitions, as well as the direct determination of the bandgap from the fit without further bias due to the overlap of the Urbach tail. In order to examine the universality of the shape of these models, we carry on a dimensionless JDOS analysis of the fundamental absorption. The importance of this scheme is the dependency of the models on a single parameter, forming a universal curve which can be used for comparison purposes. What is more, experimental results can be brought to this analysis \cite{thevaril,oleary}. The procedure consist in rewriting the aforementioned models in terms of a dimensionless independent variable $z=\beta(E-E_{g})$ \cite{thevaril,oleary,guerra}. The permittivity proportional to the JDOS is then divided by the Lorentz oscillator L(E) component along with the multiplying constants. Thus, our quantity $\varepsilon_{2}.E\sqrt{\beta}/C.L(E)$ and $\varepsilon_{2}.E\beta^2/\tilde{C}.L(E)$ will be presented as $\mathcal{D}_{cv}(z)$ and $\mathcal{J}_{cv}(z)$ for direct and indirect/amorphous, respectively.

For direct electronic transitions materials we have the Ullrich-Lorentz model
\begin{equation}
\mathcal{D}_{cv}^{\mathit{UL}}(z)= \left\{
                \begin{array}{ll}
                  \sqrt{z}  & , z \geq 1/2 \\
                  \frac{1}{\sqrt{2}}e^{(z-1/2)}   & , \ z < 1/2 \\              
                \end{array}  
                \right. 
                \label{ul-model-adim} 
\end{equation}
the BF-Lorentz,
\begin{equation}
\mathcal{D}_{cv}^{\mathit{BFL}}(z)=-\frac{\sqrt{\pi}}{2}\textrm{Li}_{1/2} \left( -e^{z} \right),
\label{bfl-dir-adim}
\end{equation}
and the Monolog-Lorentz as
\begin{equation}
\mathcal{D}_{cv}^{\mathit{ML}}(z)=\textrm{log}^{1/2}\left(1+e^{z} \right).
\label{ml-dir-adim}   
\end{equation}

Whereas, for indirect electronic transitions materials we have the O'Leary-Lorentz model
\begin{equation}
\mathcal{J}_{cv}^{\mathit{OL}}(z)= \left\{
                \begin{array}{ll}
                   \Xi(z) & , \ z \geq 1/2  \\
                  \frac{1}{\sqrt{2}}e^{\left(z-1/2\right)} Y(0) & , \ z < 1/2 \\              
                \end{array}  
                \right.  
                \label{ol-model-adim}
\end{equation}
the BF-Lorentz,
\begin{equation}
\mathcal{J}_{cv}^{\mathit{BFL}}(z)=-\frac{\pi}{4}\textrm{Li}_{2} \left( -e^{z} \right), \label{bfl-ind-adim}
\end{equation}
and the Monolog-Lorentz as
\begin{equation}
\mathcal{J}_{cv}^{\mathit{ML}}(z)=\frac{\pi}{8}\textrm{log}^{2} \left( 1+e^{z} \right). \label{ml-ind-adim}
\end{equation}
\begin{figure}
\includegraphics[scale=0.48]{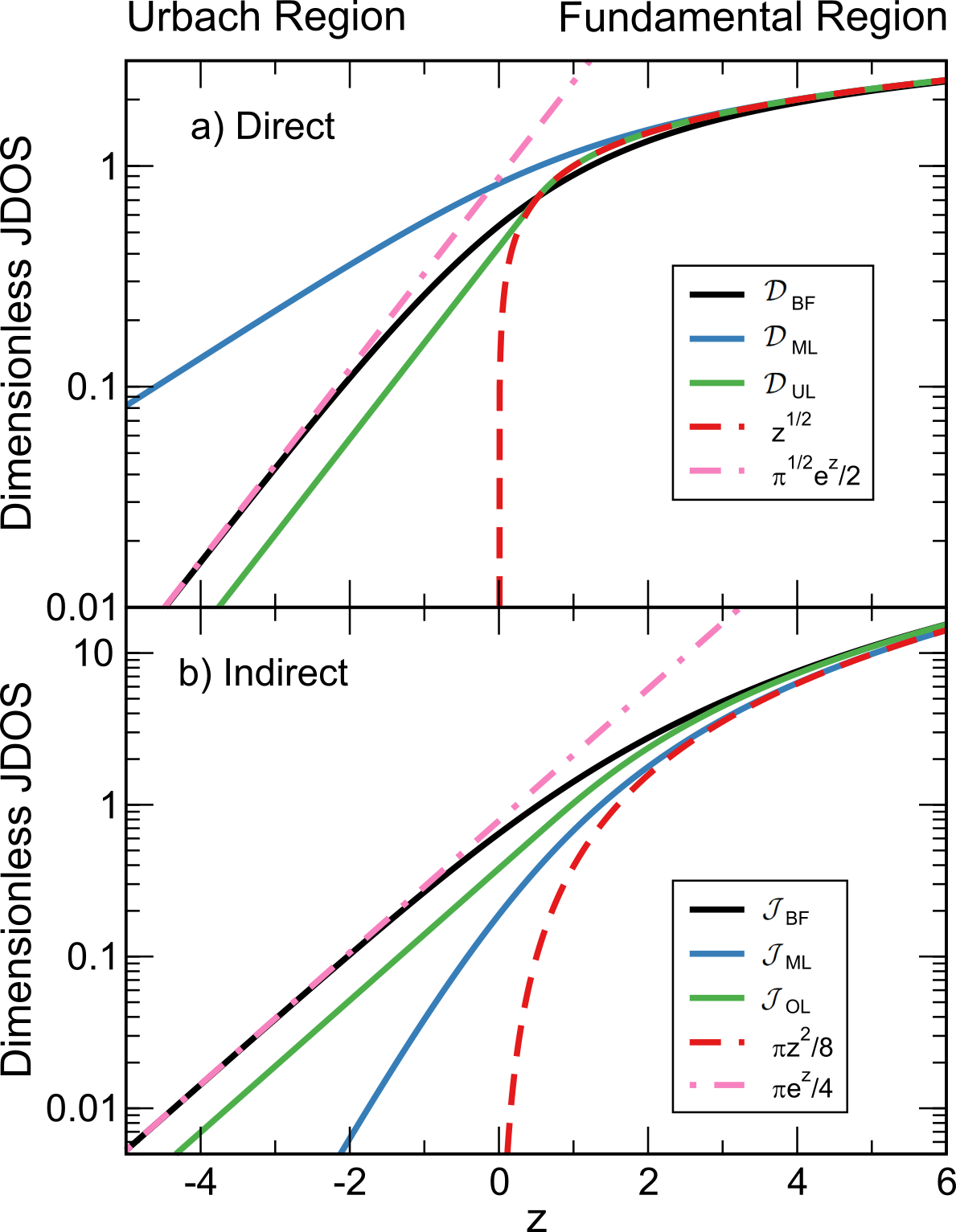}
\caption{Dimensionless JDOS for the different models. (a) Direct transition materials are represented by the models of Ullrich-Lorentz ($\mathcal{D}_{cv}^{\mathit{UL}}$), BF-Lorentz ($\mathcal{D}_{cv}^{\mathit{BFL}}$), Monolog-Lorentz ($\mathcal{D}_{cv}^{\mathit{ML}}$) and the square root behaviour. (b) Indirect transition materials are described by the models of O'Leary-Lorentz ($\mathcal{J}_{cv}^{\mathit{OL}}$), BF-Lorentz ($\mathcal{J}_{cv}^{\mathit{BFL}}$), Monolog-Lorentz ($\mathcal{J}_{cv}^{\mathit{ML}}$), and Tauc-Lorentz (quadratic behaviour). The exponential behaviour of the Urbach tail is illustrated for both cases.}
\label{fig:adim}
\end{figure}

These models in the dimensionless JDOS framework are depicted and compared in figure \ref{fig:adim}. The models share the same behaviour on the fundamental region as the BFL model. The same happens for the indirect case. On the other hand, the exponential tail exhibit different slopes depending on the model. The latter being a feature of the procedure generating the tails in each model. The ML model for Urbach region present a large (short) Urbach tail for direct (indirect) when compared to the UL (OL) model.

\section{Comparison with experiments}
Here we use the models to analyze the absorption coefficient of crystalline direct, indirect, and amorphous semiconductors. We use data of MAPI, GaAs and InP materials, to test the direct electronic transitions models. Whereas, we fit absorption coefficient data of crystalline Si and GaP and amorphous Si, to test indirect (amorphous) electronic transitions materials..

The $\varepsilon_{2}$ spectra of each type of semiconductor are analyzed as follows. First, we perform a fit of the imaginary part of the dielectric constant by using several oscillators up to a cutoff energy. In this process we have included an amount up to six oscillators in the fitting procedure. We use the BFL model to describe the fundamental absorption and TL for high absorption regions. Second, we perform an analysis exclusively of the fundamental oscillator for testing the pool of models such as UL, OL and ML. These parameters are then collected and compared for indirect and indirect, respectively. Third, we carry on a dimensionless analysis to compare the fits of different materials in the same dimensionless scale.  Fits and dimensionless JDOS analysis are presented in linear and logarithmic scales for visualization purposes only. Nevertheless, all fitting procedure was performed in linear scale. Lastly, we present the Urbach slope analysis, in which we compare the retrieved Urbach energy from Urbach's law, and the ones obtained with the here presented models.

\subsection{Direct semiconductors}

\begin{figure*}[htb]\centering 
\includegraphics[scale=0.38]{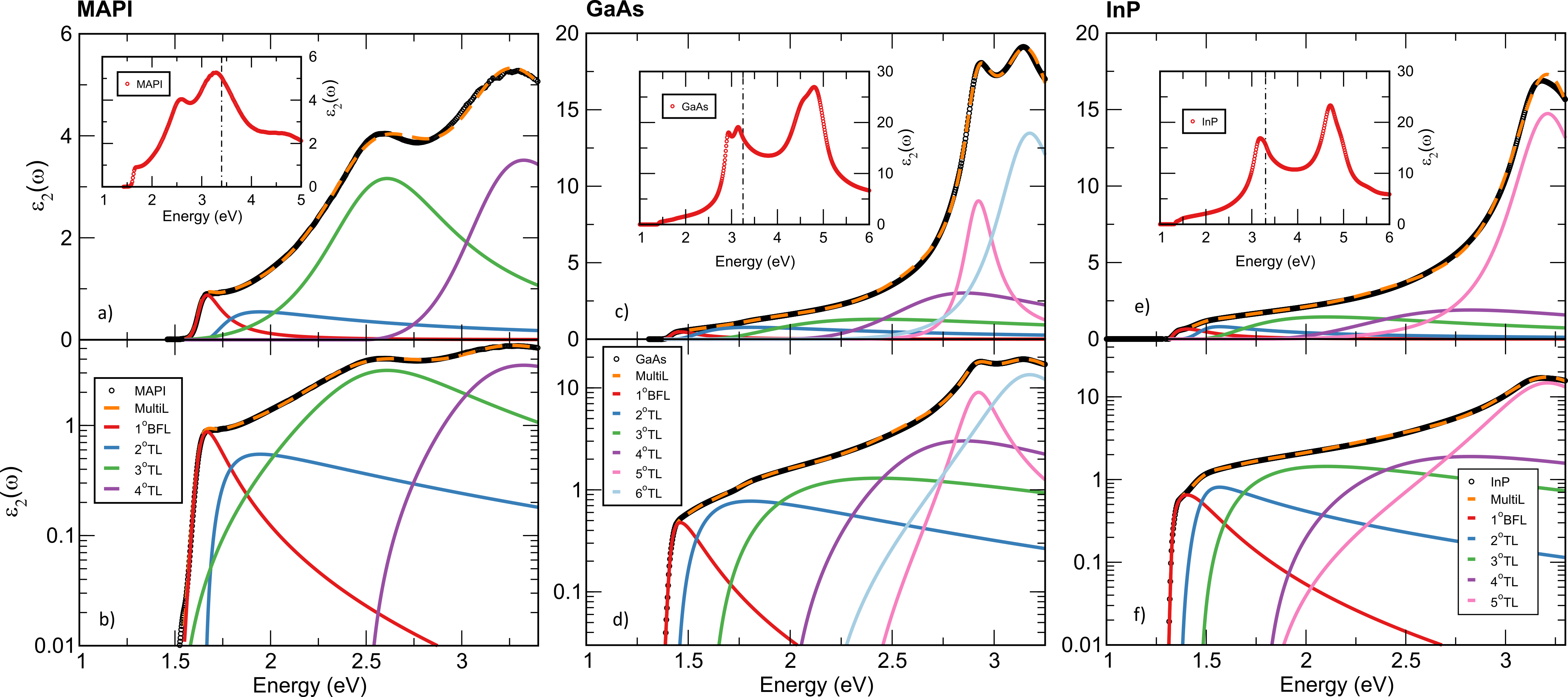}
\caption{Imaginary part of the dielectric constant of methylammonium lead iodide (MAPI) (a,b), gallium arsenide (GaAs) (c,d), and indium phosphide (InP) (e,f). The experimental data (o) and fits using multiple Lorentz oscillators (Lorentz oscillators are presented as --- ; while the sum is - - -) are illustrated on a linear and a semi-logarithmic scale. The insets in these figures show the full range of values from the VIS to UV. Here, the dashed lines represents the cutoff energy taken for the analysis.}
\label{fig:direct_multi}
\end{figure*}
\begin{figure*}[htb]\centering 
\includegraphics[scale=0.38]{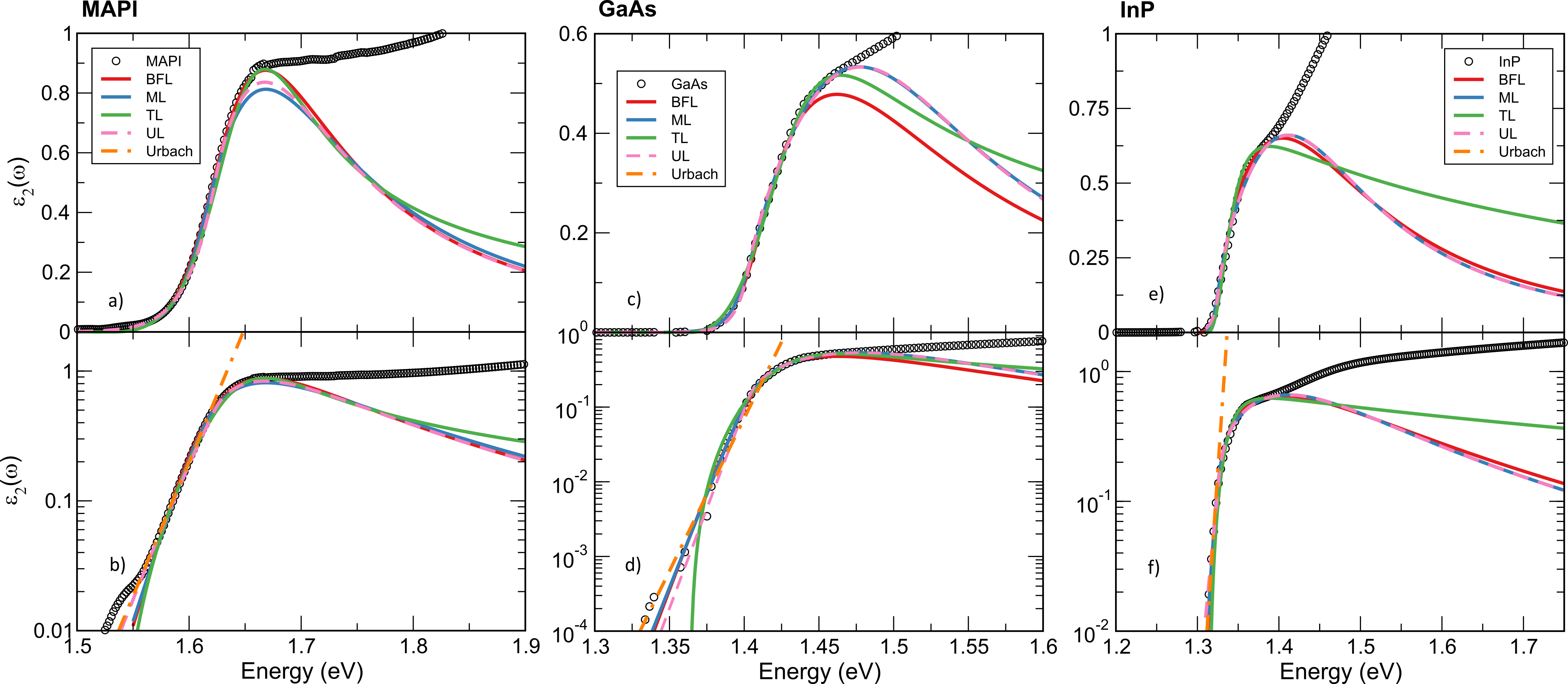}
\caption{First oscillator of the imaginary part of the dielectric constant, in normal and logarithmic scale, compared with the different theories, the BFL, ML, TL and UL for direct electronic transitions materials: MAPI (a,b), GaAs (c,d) and InP (e,f).}
\label{fig:direct_primerosc}
\end{figure*}

In the case of MAPI, the imaginary part of the dielectric constant used correspond to our previous publication, Guerra et al \cite{guerra3}, whereas the rest were obtained from the SpectraRay software libraries. The spectra are shown in figure \ref{fig:direct_multi} in normal and logarithmic scale, for viewing purposes only. The fitting procedure used consists of adding oscillators one-by-one while keeping constant values of the previous oscillator. This fixed parameters are then set free for a final fit. In every material, the first oscillator near the band-edge seen in figure \ref{fig:direct_multi} corresponds to the BFL model, whilst the added oscillators ($2^{nd}-N^{th}$) to the TL model. After this procedure, we replace the first oscillator by the corresponding to the other models for direct electronic transitions materials, i.e. UL and ML. Additionally, we use the TL model for the fundamental absorption for comparison purposes.  Fig. 4, depicts  the fits of the \ref{fig:direct_primerosc} The plots regarding the first oscillator in linear and logarithmic scales for MAPI, GaAs and InP. 

The MAPI $\varepsilon_{2}$ spectrum was fitted by using 1 BFL and 3 TL oscillators as is shown in figure \ref{fig:direct_multi}-a,b. The  oscillators parameters are shown in table \ref{table:mapi1}. The best fitted bandgap values are similar for BFL, ML and UL,  while a shift of $90$ meV is observed after the TL model. These values are in agreement with the reported of $1.60$ eV \cite{guerra2,wolf,loper} and $1.63$ eV \cite{shirayama}. We also obtain similar parameters of central energy ($E_{c}$) and damping factor ($B$) for the BFL, UL and OL models. This explains the similar curves for the high absorption region in figures $\ref{fig:direct_primerosc}$-a,b only differentiated by the coefficient $A$. Also, differences in the fitted Urbach energy  ($E_{\beta}$) values are expected, nevertheless, the fit with the BFL is the closest to previous reports \cite{ledinsky}. For comparison purposes, the central energies of the TL oscillators are also depicted in table \ref{table:mapi1}.

The GaAs spectra showed in \ref{fig:direct_multi}-c,d was analysed with 1 BFL and 5 TL oscillators. The best fitted parameters are shown in table \ref{table:GaAs1}. Notethat the bandgap is the same for BFL, ML and UL models, whilst its shifted in the TL model by $40$ meV. The bandgap values are close to previous reports of $1.422$ eV \cite{beaudoin}, $1.41$ eV \cite{ullrich3}, $1.44$ eV \cite{blakemore} and $1.43$ eV \cite{kittel}. In the case of the $1^{st}$ oscillator, the ML and UL model share almost all parameters except for the Urbach energy whose value is doubled in the case of the UL model. This  is somehow expected as depicted in the exponential slopes of figure \ref{fig:adim}. A similar behaviour is reported for BFL and UL models. Both models predict the same Urbach tail but differs in the high absorption region due to the difference of $E_{c}$ and $A$. On the other hand, the Urbach energy found in literature is of $7.5$ meV \cite{johnson}, which is in close agreement with the obtained after fitting the  BFL and UL models. For comparison purposes, the TL oscillators central energies are also presented in table \ref{table:GaAs1}.

The InP imaginary dielectric constant shown in \ref{fig:direct_multi}-e,f was fitted by using 1 BFL and 4 TL oscillators. The best fitted parameters are written in table \ref{table:InP1}. The bandgap values for the first oscillator is the same for all models, within a difference of $20$ meV, even for TL. In this case, the TL curve tries to cover the steep slope observed in figures \ref{fig:direct_primerosc}-e,f. The fitted bandgap values are close to the previously reported of $1.27$ eV \cite{kittel}, $1.343$ \cite{beaudoin}, and $1.35$ eV \cite{herzinger}, $1.37$ eV \cite{subedi}. Fits with the ML and UL models exhibit similar best fitted parameters, as in the case of GaAs, except for the Urbach energy. The Urbach energy of $6.71$ meV retrieved from the UL model is the closest to the previously reported result of $7.1$ meV \cite{beaudoin}. Ultimately, the central energies for the TL oscillators are collected in table \ref{table:InP1}.

\begin{table*}
\caption{\label{table:mapi1}MAPI parameters for the BFL, TL, ML and UL used for the fundamental oscillator (N=1). And the TL parameters for the N$^{th}$ oscillators.}
\begin{indented}
\item[]\begin{center}
\begin{tabular}{@{}cccccccc}
\br
N&\centre{4}{1st}&2nd&3rd&4th\\
\ns
&\crule{4}&\crule{3}\\
& BFL & TL & ML & UL & \centre{3}{TL} \\ \mr 
 $E_\beta$ (meV) & 17.05 & - & 9.21 & 23.80 &- &- &- \\
 $A$ & 2.79 & 18.84 &  2.67 & 2.41 &32.65 &15.51 &52.75   \\ 
 $Eg$ (eV) & 1.62 & 1.53 &  1.61 & 1.61 &1.65 &1.49 &2.48 \\ 
 $Ec$ (eV) & 1.62 & 1.64 & 1.61 & 1.62 &1.68 &2.55  &3.18 \\ 
  $B$ & 0.18 & 0.12 & 0.21 & 0.20 &0.28 &0.86 &0.81 \\ 
  \br
\end{tabular}
\end{center}
\end{indented}
\end{table*}

\begin{table*}
\caption{\label{table:GaAs1}GaAs parameters for the BFL, TL, ML and UL used for the fundamental oscillator (N=1). And the TL parameters for the N$^{th}$ oscillators.}
\begin{indented}
\item[]\begin{center}
\begin{tabular}{@{}cccccccccc}
\br
N&\centre{4}{1st}&2nd&3rd&4th&5th&6th\\
\ns
&\crule{4}&\crule{5}\\
& BFL & TL & ML & UL & \centre{5}{TL} \\ \mr 
 $E_\beta$ (meV) & 8.62 & - & 4.60 & 8.62 &- &- &- &- &-\\
 $A$ & 1.23 & 19.80 & 1.07 & 1.00 & 35.24 & 35.88 & 36.08 &39.06 &52.22 \\ 
 $Eg$ (eV) & 1.41 & 1.36  &  1.40  & 1.40  & 1.43 &1.60 &1.96  &2.31 &2.12 \\ 
 $Ec$ (eV) & 1.41   & 1.42 & 1.44  & 1.45  &1.43 &1.76 &2.68 &2.91  &3.15 \\ 
  $B$ & 0.22  & 0.11 & 0.23 & 0.22 &0.38 &1.30 &0.98 &0.19 &0.42 \\ 
  \br
\end{tabular}
\end{center}
\end{indented}
\end{table*}
\begin{table*}
\caption{\label{table:InP1}InP parameters for the BFL, TL, ML and UL used for the fundamental oscillator (N=1). And the TL parameters for the N$^{th}$ oscillators.}
\begin{indented}
\item[]\begin{center}
\begin{tabular}{@{}ccccccccc}
\br
N&\centre{4}{1st}&2nd&3rd&4th&5th\\
\ns
&\crule{4}&\crule{4}\\
& BFL & TL & ML & UL & \centre{4}{TL} \\ \mr 
 $E_\beta$ (meV) & 4.34 & - &    2.31 & 6.71 &- &- &- &-\\ 
 $A$ & 1.66 & 69.99 &  1.38 & 1.37 &23.30 &42.67 &29.24 &35.13\\ 
 $Eg$ (eV)& 1.33 & 1.31 &  1.32 & 1.32 &1.37 &1.46 &1.78 &1.77\\
 $Ec$ (eV)& 1.35 & 1.32 & 1.37 & 1.37 &1.45 &1.53 &2.46 &3.20\\ 
  $B$ & 0.31 & 0.06 & 0.28 & 0.28 &0.23 &0.90 &1.53 &0.42\\
  \br
\end{tabular}
\end{center}
\end{indented}
\end{table*}

\subsection{Indirect Semiconductors}
\begin{figure*}[ht]\centering 
\includegraphics[scale=0.4]{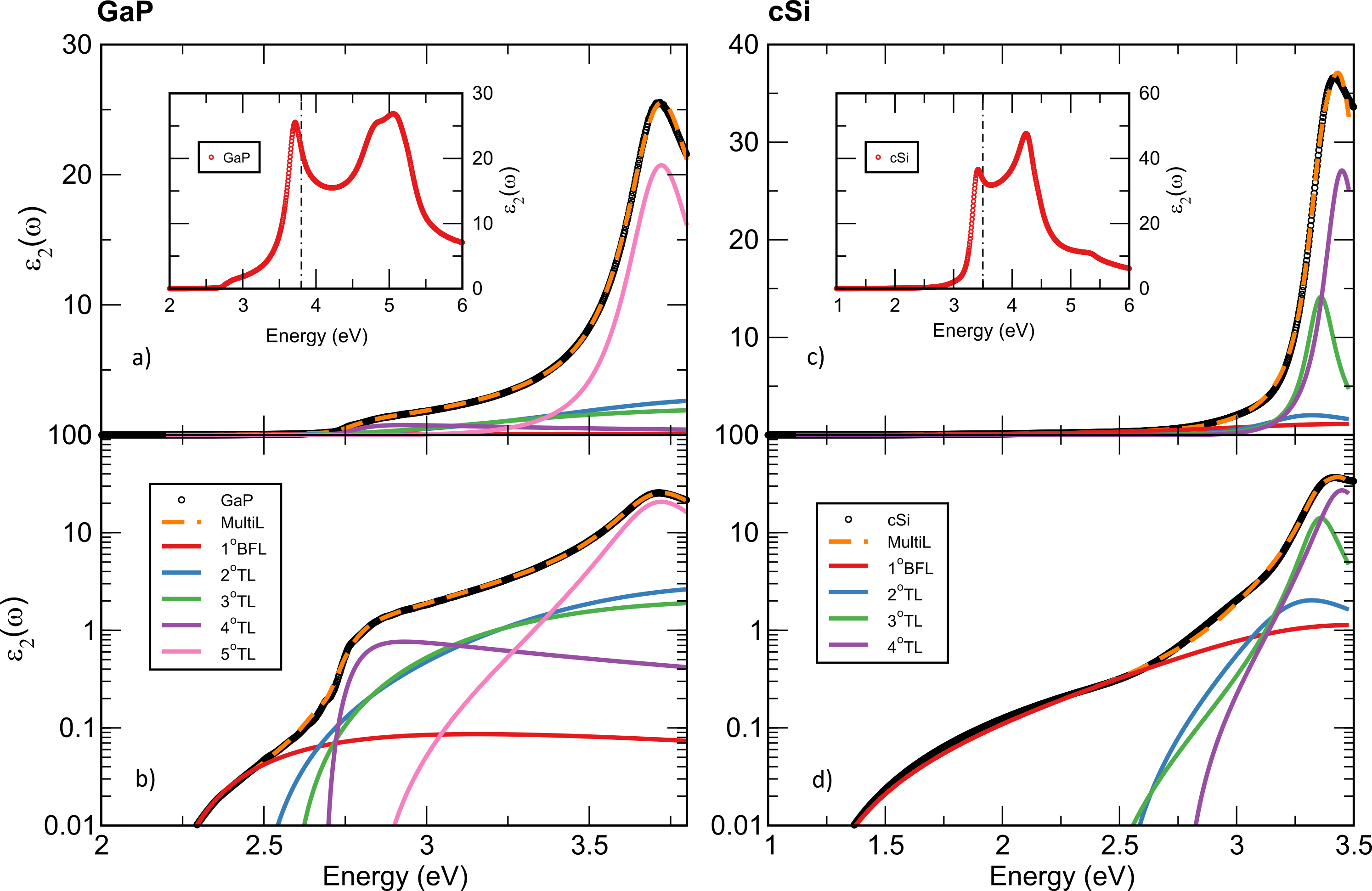}
\caption{Imaginary part of the dielectric constant of gallium phosphide (GaP) (a,b) and cristalline silicon (cSi) (c,d). The experimental data (o) and fits using multiple Lorentz oscillators (Lorentz oscillators are presented as --- ; while the sum is - - -) are illustrated on a linear and a semi-logarithmic scale. The insets in these figures show the full range of values from the VIS to UV. Here, the dashed lines represents the cutoff energy taken for the analysis.}
\label{fig:indir_multi}
\end{figure*}

\begin{figure*}[ht]\centering 
\includegraphics[scale=0.38]{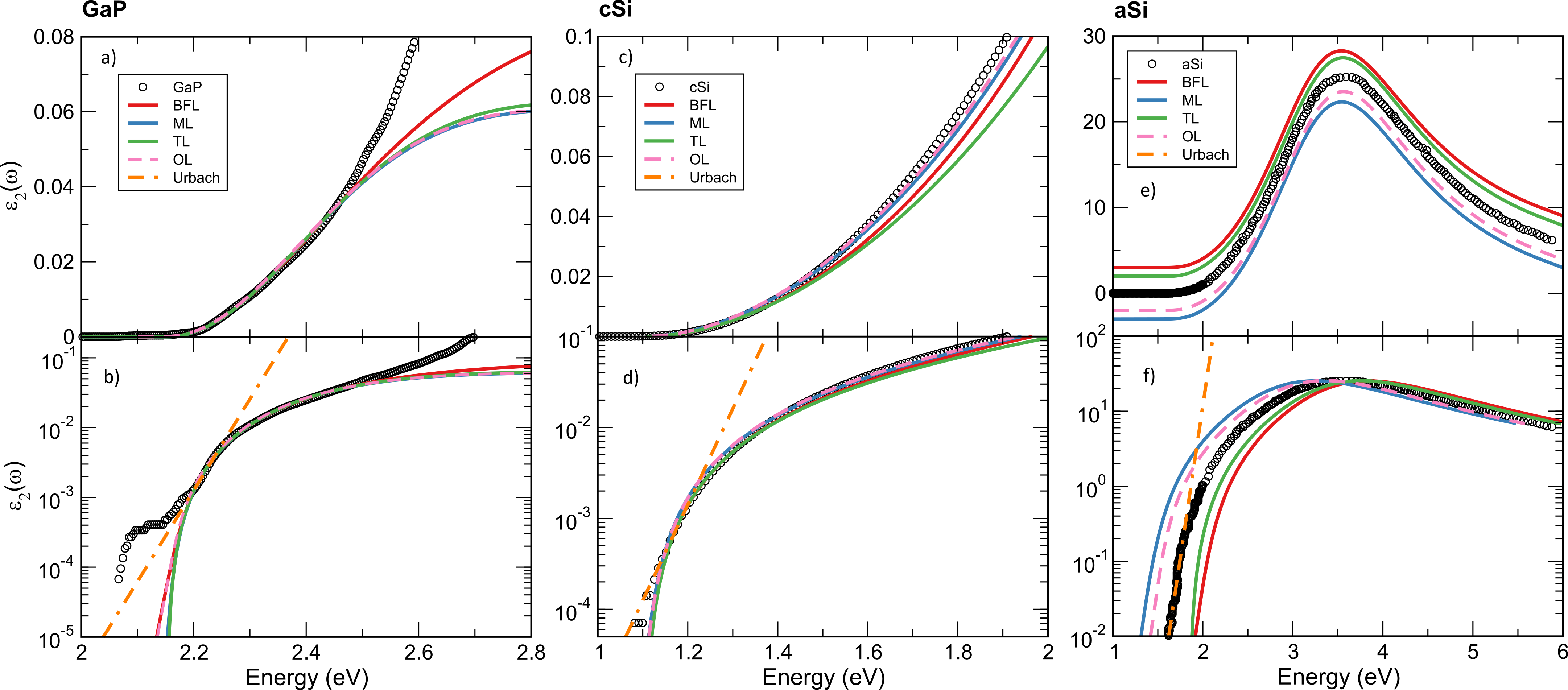}
\caption{First oscillator of the imaginary part of the dielectric constant, in normal and logarithmic scale, compared with the different theories, the BFL, ML, TL and UL for direct electronic transitions materials: GaP (a,b), cSi (c,d) and aSi (e,f). Notice that for clarity purposes, the dielectric constant of the models is shifted in (e) by an amount of $\pm3$ and $\pm2$ units. This is repeated for the energy in (f) with amounts of $\pm0.3$eV and $\pm0.2$eV.}
\label{fig:indir_primerosc}
\end{figure*}
The imaginary part of their dielectric constant was extracted from SpectraRay libraries.  GaP  $\varepsilon_2$ data was extracted from Aspnes et al. \cite{aspnes}, and the corresponding for c-Si was selected from the UV-NIR data. These are shown in figure \ref{fig:indir_multi} in linear and logarithmic scale. The fitting procedure employed was the same as for direct semiconductors. We have used the BFL model for fundamental absorption and $N^{th}$ TL oscillators for higher electronic transition  energies. The cutoff energy is depicted in  the inset of figures \ref{fig:indir_multi}-a,c, for each case. Fits of the first oscillator using the BFL, OL and ML models are shown in figure \ref{fig:indir_primerosc}-a,b,c,d for linear and logarithmic scale. 

GaP $\varepsilon_2$ spectrum was fitted using 1 BFL and 4 TL oscillators as shown in figure \ref{fig:indir_multi}-a,b. The fundamental absorption is only seen in the logarithmic scale due to the small components of the transition matrix element for the phonon assisted transition. Best fitted parameters  are written in table \ref{table:GaP1}. Bandgap values are close between each model and differ about $100$ meV with respect to the $2.25$ eV reported in \cite{kittel} at room temperature. The first oscillator exhibit virtually the same best fitted parameters between the ML and OL models. Figure \ref{fig:indir_primerosc}-a,b depicts the aforementioned fits.

c-Si $\varepsilon_2$ data  was fitted with 1 BFL and 3TL oscillators. This is shown in figure \ref{fig:indir_multi}-c,d. The best fitted parameters are written in table \ref{table:cSi1}. The small fundamental indirect absorption is only visible in logarithmic scale as in the case of GaP. The retrieved bandgap value of $1.10$ eV is the same between all models. This is in agreement with the well known value of $1.11$ eV \cite{kittel}, and $1.124$ eV calculated with the free-exciton absorption \cite{bludau}. The Urbach tail calculated with BFL/ML/OL gives a value of $10$ meV which differs from the literature value of $45$ meV of \cite{cody}. This apparent difference can be attributed to the sample conditions and preparation.

\begin{table*}
\caption{\label{table:GaP1}GaP parameters for the BFL, TL, ML and OL used for the fundamental oscillator (N=1). And the TL parameters for the N$^{th}$ oscillators.}
\begin{indented}
\item[]\begin{center}
\begin{tabular}{@{}ccccccccc}
\br
N&\centre{4}{1st}&2nd&3rd&4th&5th\\
\ns
&\crule{4}&\crule{4}\\
& BFL & TL & ML & OL & \centre{4}{TL} \\ \mr 
 $E_\beta$ (meV) & 10 & - & 10 & 10 &- &- &- &- \\ 
 $A$ & 11.07 & 2.45 &  6.05 & 6.05 &40.08 &61.77 &95.49 &95.49 \\ 
 $Eg$ (eV) & 2.16 & 2.15 &  2.15 & 2.15 &2.46 &2.57 &2.68 &2.81 \\
 $Ec$ (eV) & 2.16 & 2.32 & 2.32 & 2.32 &3.76 &3.26 &2.74 &3.71 \\ 
  $B$ & 1.35 & 0.94 & 0.91 & 0.91 &1.88 &2.52 &0.24 &0.27 \\
  \br
\end{tabular}
\end{center}
\end{indented}
\end{table*}

\begin{table*}
\caption{\label{table:cSi1}c-Si parameters for the BFL, TL, ML and OL used for the fundamental oscillator (N=1). And the TL parameters for the N$^{th}$ oscillators.}
\begin{indented}
\item[]\begin{center}
\begin{tabular}{@{}cccccccc}
\br
N&\centre{4}{1st}&2nd&3rd&4th\\
\ns
&\crule{4}&\crule{3}\\
& BFL & TL & ML & OL & \centre{3}{TL} \\ \mr 
 $E_\beta$ (meV) & 10 & - & 10 & 10 &- &- &-\\ 
 $A$ & 9.64 & 9.58 & 20.50 & 20.52 &17.68 &24.35 &136.92\\ 
 $Eg$ (eV)& 1.10 & 1.10 &  1.10 & 1.10 &2.47 &2.35 &2.76\\ 
 $Ec$ (eV)& 3.48 & 4.20 & 3.88 & 3.98 &3.26 &3.36 &3.44\\ 
  $B$ & 1.57 & 1.14 & 1.17 & 1.34 &0.55 &0.16 &0.20\\ 
  \br
\end{tabular}
\end{center}
\end{indented}
\end{table*}

\subsection{Amorphous Materials}
The fitting procedure used for Amorphous materials is the same as for indirect electronic transitions materials. Here we analyze a-Si. Absorption data was extracted from Jackson et al. \cite{jackson}. This is shown in figure \ref{fig:indir_primerosc}-e,f. Note the presence of a single oscillator. The best fitted parameters are collected in table \ref{table:aSi1}. The same bandgap energy is obtained for all models. The reported bandgap values of a-Si of $1.72$ eV \cite{cody,cody1981} confirms the good estimation of our results. From the functional behaviour (see figure \ref{fig:indir_primerosc}-e,f), the central energy $E_{c}$ and broadening factor $B$ have similar values for all models. The Urbach energy of BFL and OL models are the closest to Cody's value of $45$ meV \cite{cody}.
\begin{table}
\caption{\label{table:aSi1}a-Si parameters for the BFL, TL, ML and OL used for its single oscillator, the fundamental N=1.}
\begin{indented}
\item[]\begin{center}
\begin{tabular}{@{}ccccc}
\br
N&\centre{4}{1st}\\
\ns
&\crule{4}\\
& BFL & TL & ML & OL  \\ \mr 
 $E_U$ (meV)& 47.55 & - & 80.51 & 52.57  \\ 
 $A$ & 525.78 & 194.47 & 516.43 & 501.41\\ 
 $Eg$ (eV& 1.68 & 1.65 & 1.67 & 1.67\\ 
 $Ec$ (eV)& 3.39 & 3.44 & 3.41 & 3.44\\ 
  $B$ & 2.13 & 2.11 &  2.12 & 2.10 \\ 
  \br
\end{tabular}
\end{center}
\end{indented}
\end{table}

\subsection{Dimensionless comparison}

\begin{figure*}[ht]\centering 
\includegraphics[scale=0.35]{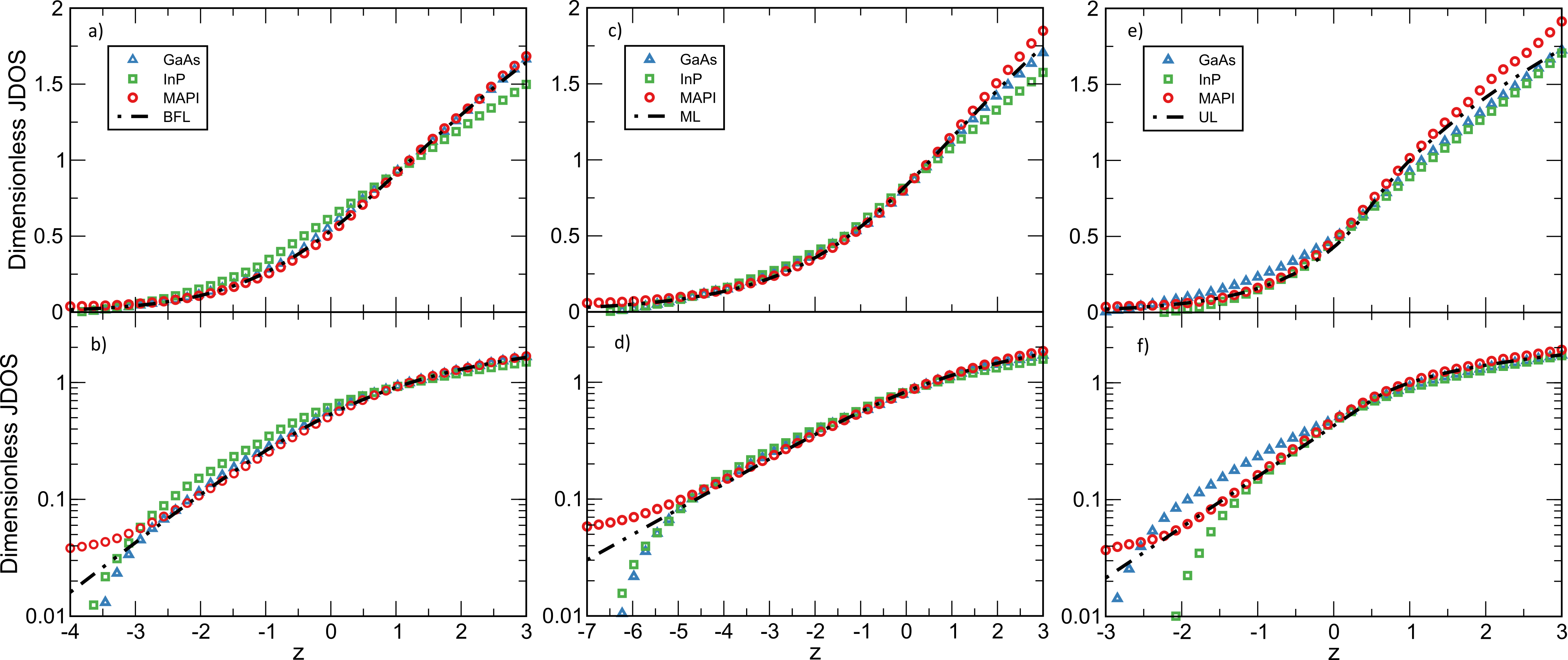}
\caption{Dimensionless JDOS ($\mathcal{D}_{cv}$) for direct electronic transitions materials (GaAs, InP and MAPI), in normal and logarithmic scale, compared with the dimensionless models such as BF-Lorentz (a,b), Monolog-Lorentz (c,d) and Ullrich-Lorentz (e,f).}
\label{fig:adim_direct_join}
\end{figure*}

\begin{figure*}[ht]\centering 
\includegraphics[scale=0.35]{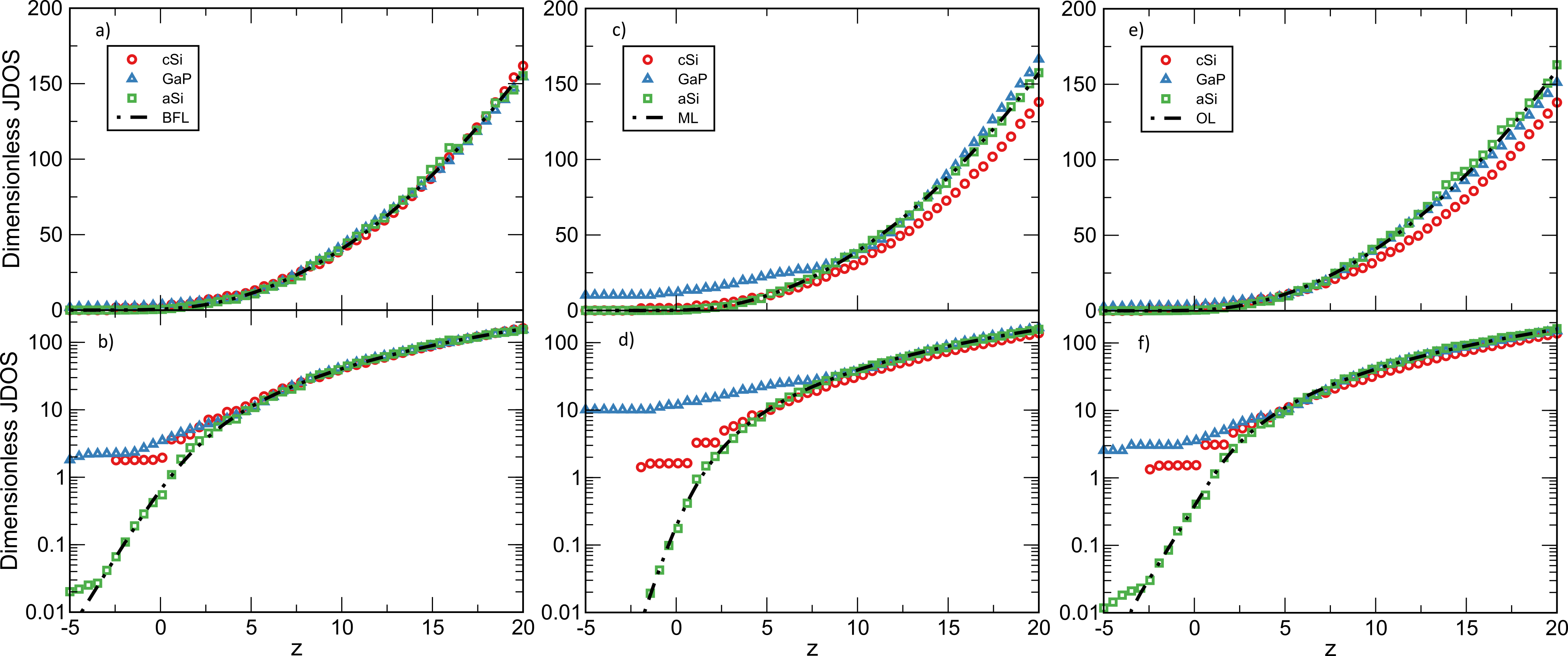}
\caption{Dimensionless JDOS ($\mathcal{J}_{cv}$) for indirect electronic transitions materials (GaP and c-Si) and amorphous Si, in normal and logarithmic scale, compared with the dimensionless models such as BF-Lorentz  (a,b), Monolog-Lorentz (c,d) and O'Leary-Lorentz (e,f).}
\label{fig:adim_indirect_join}
\end{figure*}

We now carry on a dimensionless analysis for direct, indirect (amorphous) materials. This is presented in figures \ref{fig:adim_direct_join} and \ref{fig:adim_indirect_join}, respectively. The objective of this analysis is to contrast the different models, in the same dimensionless scale. Figures \ref{fig:adim_direct_join} and \ref{fig:adim_indirect_join} depict the dimensionless JDOS  curves along with the spectral data of each direct and indirect (amorphous) material brought into this scale, respectively. Differences observed are attributed to the model goodness. In the particular case of c-Si and c-GaP the Urbach tail region data could be close to the spectral sensitivity of the instrument.

\subsection{Urbach slope}
\begin{figure*}[htb]\centering 
\includegraphics[scale=0.5]{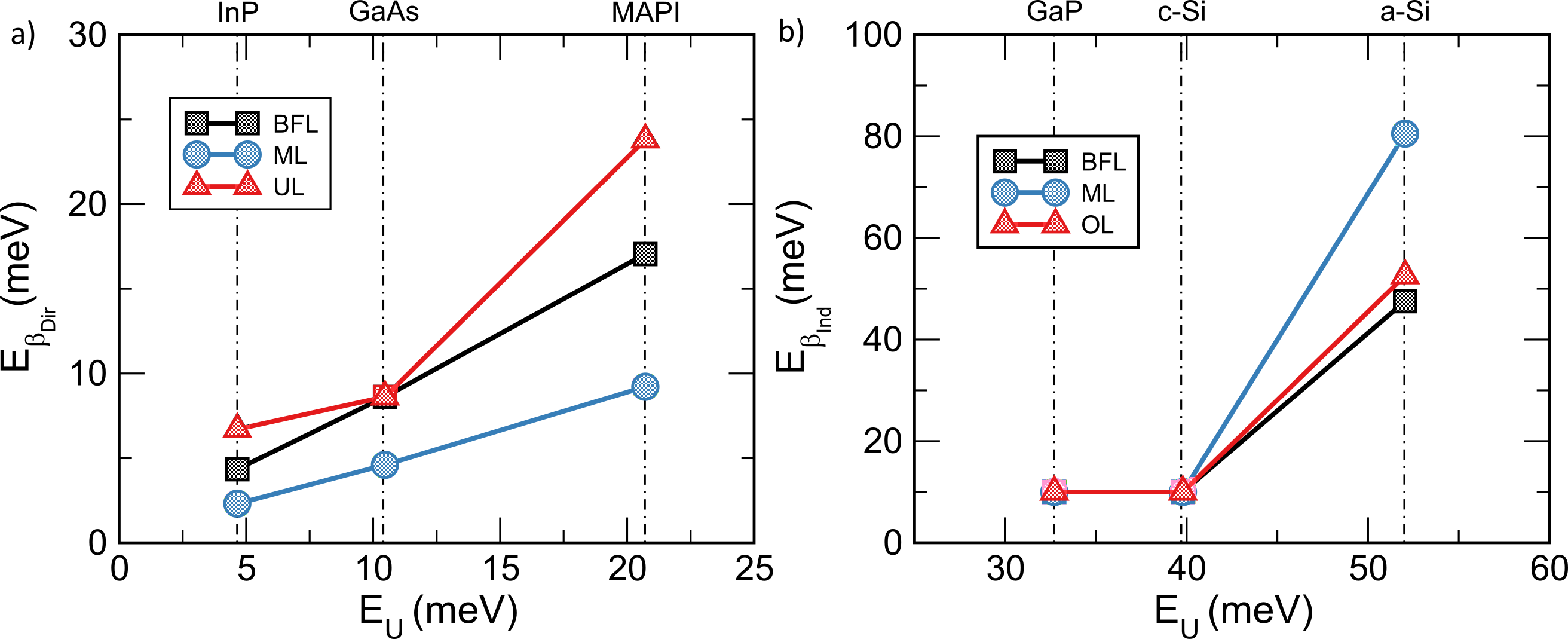}
\caption{Urbach energies calculated with BFL/ML/UL(OL) models plotted with the traditional Urbach energy extracted from the Urbach's rule (eq. \ref{urbach_tail}). The case of direct electronic transitions materials is developed in (a) and indirect/amorphous are shown in (b).}
\label{fig:beta_urbach}
\end{figure*}

In this subsection we analyze in detail the Urbach slope of the different models and compare them with the Urbach tail obtained with the traditional Urbach rule model. The comparison between the Urbach energy for direct $E_{\beta_{dir}}$ and indirect $E_{\beta_{ind}}$ semiconductors with the Urbach energy from the Urbach rule ($E_U$) is shown in figure \ref{fig:beta_urbach},  respectively. In both cases we see that Urbach energies depicted with the models are shifted when compared to $E_U$. This shift is larger for larger Urbach energies. The $E_U$ for the direct MAPI, GaAs and InP is $20.7$ meV, $10.47$ meV and $4.7$ meV, respectively. These values are in agreement with other results using the Urbach rule. For instance, the literature reports are $14$ meV \cite{ledinsky} for MAPI, $7.5$ meV \cite{johnson} for GaAs, and $7.1$ meV \cite{beaudoin} for InP. 

From figure \ref{fig:beta_urbach}-a, it can be noticed that the BFL and ML models tend to behave  linearly when compared with $E_U$. This supports the correct function-ability of the PolyLog in the Urbach region. On the other hand, the Urbach energies for the UL model are similar to the BFL for InP and GaAs, except for MAPI. Lastly, the disorder energy of ML model is  half the value derived from the Urbach rule for all materials. 

In the case of indirect/amorphous materials (see figure \ref{fig:beta_urbach}-b), the Urbach energies computed with the Urbach rule are $52$ meV, $39.7$ meV and $32.7$ meV for a-Si, c-Si and GaP, respectively. The corresponding literature values are $45$ meV and $12$ meV for a-Si and c-Si, respectively \cite{cody,macfarlane}, but these can vary depending on the growing process, defects concentration and deposition temperature \cite{falsini,azmi}. The difference in Urbach energies for c-Si may be due to experimental sensitivity when compared with MacFarlane's result \cite{macfarlane}. Despite this, MacFarlane's  value is similar to the $10$ meV value of $E_{\beta_{ind}}$ depicted with our BFL/OL/ML models. We can conclude that for c-Si and GaP, the $E_{\beta_{ind}}$ is the same for the BFL, OL and ML models. In the case of GaP, its value is the third part of the traditional Urbach energy. Lastly, in the case of the large tail of a-Si, the BFL and OL models are the closest to the value of $E_U$. While the ML differs by an amount of $30$ meV. 

\section{Conclusion}

We have presented a review of the models developed for the accurate description of different regions of the absorption coefficient. An adequate model for describing the fundamental and high absorption regions of direct semiconductors taking into account the Urbach tail is missing. Tauc-Lorentz and Cody-Lorentz models serve as inspiration to develop new self consistent models. After following the  procedure of Jellison-Modine, we arrive to our versions of Ullrich-Lorentz (UL), O'Leary-Lorentz (OL), Band-Fluctuations-Lorentz (BFL) and Monolog-Lorentz (ML) for direct and indirect/amorphous semiconductors. Their advantages are the incorporation of the Urbach tail, the smooth transition from fundamental to high absorption and the dependency on 5 fitting parameters only. What is more, the BFL model describes direct, indirect and amorphous semiconductors within a theory that arises from the same principles, while the ML model overcomes the difficulties of the BFL by producing an analytic equation.

We have tested our models for direct (MAPI, GaAs, InP), indirect (c-Si, GaP) and amorphous (a-Si) materials with excellent agreement between experiment and models. Our analysis has been done by fitting several oscillators for the whole spectra up to a cutoff energy. We have also analyzed the first oscillator describing the fundamental absorption for each model. These results have also been set in the dimensionless framework for comparison purposes. This extended procedure has been helpful for assessing the capabilities of each model to describe properly the fundamental absorption region along with the high absorption part. The values obtained for each model are in good agreement with the results found in the literature. We believe these models will be helpful for experimentalists studying these and other materials.

\section*{Acknowledgments}
We would gratefully like to acknowledge the Peruvian National Council for Science, Technology and Technological Innovation (CONCYTEC) for a Ph.D. scholarship under grant no. 236-2015-FONDECYT, the German Academic Exchange Service (DAAD) in conjunction with FONDECYT (grants 57508544 and 423-2019-FONDECYT, respectively), the Helmholtz Association for funding within the HySPRINT Innovation lab project, as well as the Office of Naval Research, Grant No. N62909-21-1-2034.

\section*{References}
\bibliography{iopart-num}

\end{document}